\def\tr{{\rm tr}\,}
\def\Tr{{\rm Tr}\,}
\def\sgn{{\rm sgn\,}}
\def\b{\bibitem}
\documentstyle[aps,prb,eqsecnum,psfig,floats]{revtex}
\begin{document}
\def\SNG{{\em Physical Review Style and Notation Guide}}
\def\LUG {{\em \LaTeX{} User's Guide \& Reference Manual}}
\def\btt#1{{\tt$\backslash$\string#1}}%
\def\REVTeX{REV\TeX}
\def\AmS{{\protect\the\textfont2
        A\kern-.1667em\lower.5ex\hbox{M}\kern-.125emS}}
\def\AmSLaTeX{\AmS-\LaTeX}
\def\BibTeX{\rm B{\sc ib}\TeX}
\twocolumn[\hsize\textwidth\columnwidth\hsize\csname@twocolumnfalse%
\endcsname
 
\title{Theory of many-fermion systems II:
       The case of Coulomb interactions \\
      \small{$[$ Phys. Rev. B {\bf 58}, 9710 (1998) $]$}
}
\author{D.Belitz and F.Evers\cite{byline}}
\address{Department of Physics and Materials Science Institute\\
University of Oregon,\\
Eugene, OR 97403}
\author{T.R.Kirkpatrick}
\address{Institute for Physical Science and Technology, and Department of Physics\\
University of Maryland,\\ 
College Park, MD 20742}

\date{\today}
\maketitle

\begin{abstract}
In a recent paper a general field-theoretical description of 
many-fermion systems with short-ranged interactions has been developed.
Here we extend this theory to the case of disordered electrons interacting
via a Coulomb potential. A detailed discussion is given of the Ward identity
that controls the soft modes in the system, and the generalized nonlinear
$\sigma$ model for the Coulombic case is derived and discussed.

\end{abstract}
]

\section{Introduction}
\label{sec:I}

The theory of many-fermion systems is central to our understanding of
condensed matter, as well as of nuclei and certain astrophysical systems.
Of particular interest in a condensed matter context are the universal 
properties of many-electron systems,
i.e. phenomena at long wavelength and small frequency scales that are
independent of the material's detailed microscopic structure. Historically,
there have been two important avenues to this problem: Landau's
phenomenological Fermi-liquid theory,\cite{BaymPethick}
and the microscopic many-body perturbation theory or Feynman diagram 
approach.\cite{FetterWalecka} The latter soon was
generalized to include the scattering of electrons by 
static impurities,\cite{AGD}
and it has led to many important insights concerning the nature of
interacting disordered electron systems. For instance, it was used to
show that the combined effects of disorder and interactions lead to
the nonanalyticities in the frequency and wavenumber
dependence of both thermodynamic and transport properties of disordered
metals that have become known as `weak localization 
effects'.\cite{AAKL,WeakLocalizationFootnote}

There are, however, many
interesting phenomena for which an approach based on many-body
perturbation theory is not feasible. An example is the Anderson-Mott
metal-insulator transition that the electrons undergo with increasing
disorder strength.\cite{R} This quantum phase transition is best studied by
means of effective field theories and the use of the renormalization
group, an approach that was pioneered by Wegner.\cite{Wegner} Also, many-body
perturbation theory is ultimately unsatisfactory as a tool even in the
metallic phase, despite its impressive successes. One major problem is that,
within perturbation theory, it is not clear whether the weak-localization
effects are actually the leading nonanalyticities.
Furthermore, universal phenomena like the
weak-localization effects arise from the presence of soft modes in the problem,
which many-body perturbation theory is not well suited to deal with.
The softness or masslessness of these modes is the result of symmetries that
are often not explicit in the usual perturbative formalism, and therefore soft
modes arise apparently accidentally as a result of complicated cancellations,
rather than being manifestly built into the formalism. Finally, it is
unsatisfactory to have entirely different approaches to the universal 
properties of the various phases on the one hand, and to those of the
transitions between these phases on the other hand.

The concepts for overcoming these problems exist; they consist of a systematic
application of renormalization group ideas. Historically, the renormalization
group has been mainly associated with critical phenomena at continuous phase
transitions, for which the importance of symmetries and soft modes has been
universally appreciated. While it has always been clear in
principle, and occasionally has been emphasized,\cite{AndersonFisher}
that the renormalization group, rather than just being a tool for studying
critical phenomena, allows for a unified description of both phases and
phase transitions, these ideas have not been widely appreciated. Only
very recently has there been the beginning of a paradigm shift in this
respect. For instance, starting with Shankar's work,\cite{Shankar}
there has been much activity recently on the derivation of clean Fermi-liquid
theory as a stable renormalization group fixed point.\cite{RG_FL} 
These methods, however, have proven very hard to generalize to the 
case of quenched disorder.

In a previous paper, to be referred to as I,\cite{fermions} 
two of us have developed 
an effective field theory for many-electron systems that is particularly 
suitable for dealing with disordered systems. This theory, which is
formulated in terms of classical matrix fields, 
allows for a systematic separation
of soft and massive modes, and the latter can be integrated out in a
simple approximation to yield an effective theory for the soft modes.
The theory also allows for the clean limit to be taken, but the soft-mode
structure in that case turns out to be more complex, which gives the
effective theory fewer advantages over traditional approaches than is
the case in the presence of disorder. This theory has been used,
{\it inter alia}, to prove that the well-known weak-localization effects
are indeed the leading nonanalyties, to provide a technically satisfactory
derivation of Finkel'stein's generalization\cite{Finkelstein} of 
Wegner's nonlinear $\sigma$ model,\cite{Wegner} and to derive a 
previously unknown nonanalyticity in the spin
susceptibility of clean Fermi liquids.\cite{chi_s,fermions}

In I, the effective classical field theory was developed for fermions
that interact via a short-ranged potential, assuming that the underlying
Coulomb potential had been screened at the level of the basic fermionic
theory. In the present paper, we show how to handle a long-ranged interaction
entirely within the framework of the classical matrix field theory,
focussing on the disordered case. We 
give a detailed discussion of the Ward identity that controls the structure
of the soft modes, and we derive Finkel'stein's generalized nonlinear
$\sigma$ model\cite{Finkelstein} for the Coulombic case.

The outline of our paper is as follows. In Sec.\ \ref{sec:II} we recall the
matrix field theory of I, and slightly generalize it to allow for a
Coulomb interaction. We show that the saddle point considered in I remains
valid, and that an expansion about the saddle point to Gaussian order
produces RPA-type screening. In Sec.\ \ref{sec:III} we perform a symmetry
analysis, and derive and discuss the Ward identity that controls the soft 
modes in the
system. In Sec.\ \ref{sec:IV} we show that integrating out the massive modes
in the simplest approximation that respects the Ward identity leads to
Finkel'stein's model. In Sec.\ \ref{sec:V} we conclude by discussing our 
results, and give in particular a discussion of the accuracy of the $\sigma$
model.

\section{Matrix field theory}
\label{sec:II}

\subsection{Grassmannian field theory}
\label{subsec:II.A}

Our model and basic theoretical setup is the same as in I. We will therefore
restrict ourselves to discussing the changes that are necessary to accomodate
a Coulomb interaction.

The action is given by
\begin{mathletters}
\label{eqs:2.1}
\begin{equation}
S = -\int dx\sum_{\sigma}\ \bar{\psi}_{\sigma}(x)\,\partial_{\tau}\,
\psi_{\sigma}(x)\ + S_0 + S_{\rm dis} + S_{\rm int}\quad. 
\label{eq:2.1a}
\end{equation}
Here the $\bar\psi$ and $\psi$ are Grassmann valued fields, and
we use a ($d+1$)-vector notation, with $x=({\bf x},\tau)$, and
$\int dx=\int_V d{\bf x}\int_{0}^{\beta} d\tau$. ${\bf x}$ denotes position,
$\tau$ imaginary time, $V$ is the system volume, $\beta =1/T$ is the inverse 
temperature, $\sigma$ is the spin label, and we use units such that 
$\hbar = k_B = e^2 = 1$. $S_0$ (together with the time derivative term) 
and $S_{\rm dis}$ describe free fermions, and their
interaction with a static random potential, respectively, and have been
defined in I. The random potential we assume to be Gaussian distributed,
and we employ the replica trick to handle it.
S$_{\rm int}$ describes a Coulomb interaction, which in the replicated
theory (which we denote by a tilde) takes the form
\begin{eqnarray}
{\tilde S}_{\rm int} &=& \sum_{\alpha} {\tilde S}_{\rm int}^{\alpha} = 
                 -\frac{1}{2}\sum_{\alpha}\int dx_1\,dx_2\ 
                     \sum_{\sigma_1,\sigma_2}\ v({\bf x}_1 - {\bf x}_2)
\nonumber\\
          &&\times\delta(\tau_1 - \tau_2)\,\bar{\psi}_{\sigma_1}^{\alpha}(x_1)\,
                   \bar{\psi}_{\sigma_2}^{\alpha}(x_2)\,
          \psi_{\sigma_2}^{\alpha}(x_2)\,\psi_{\sigma_1}^{\alpha}(x_1)\quad. 
\nonumber\\
\label{eq:2.1b}
\end{eqnarray}
Here $\alpha$ is the replica index, and
\begin{equation}
v({\bf x}) = 1/\vert{\bf x}\vert\quad,
\label{eq:2.1c}
\end{equation}
is the Coulomb potential. For the dimensionalities of interest, $d=2,3$,
its Fourier transform is
\begin{equation}
v({\bf q}) = (1-\delta_{{\bf q},0})\,\frac{2^{d-1}\pi}{\vert{\bf q}\vert^{d-1}}
                                         \quad,
\label{eq:2.1d}
\end{equation}
\end{mathletters}%
where the factor $1-\delta_{{\bf q},0}$ represents a uniform positive
background charge that ensures charge neutrality.

We now introduce a momentum cutoff $\lambda$, and rewrite the interacting
part of the action as
\begin{mathletters}
\label{eqs:2.2}
\begin{equation}
\tilde{S}_{\rm int}^{\,\alpha} = \tilde{S}_{\rm int}^{\,\alpha\,(1)} + 
  \tilde{S}_{\rm int}^{\,\alpha\,(2)}+\tilde{S}_{\rm int}^{\,\alpha\,(3)}\quad , 
\label{eq:2.2a}
\end{equation}
where,
\begin{eqnarray}
\tilde{S}_{\rm int}^{\,\alpha\,(1)}&=&-\frac{T}{2}\sum_{\sigma_1\sigma_2}
     \sum_{k,p} {\sum_q}^{\,\prime}
v({\bf q})\ \bar{\psi}_{\sigma_1}^\alpha (k)\,\bar{\psi}_{\sigma_2}^\alpha
                                                                       (p+q)\,
\nonumber\\
      &&\times\psi_{\sigma_2}^\alpha(p)\,\psi_{\sigma_1}^\alpha (k+q)\quad, 
\label{eq:2.2b}
\end{eqnarray}
\begin{eqnarray}
\tilde{S}_{\rm int}^{\,\alpha\,(2)}&=&-\frac T2\sum_{\sigma _1\sigma
_2}\sum\limits_{k,p}{\sum_q}^{\,\prime}
v({\bf p}-{\bf k})\,\Theta (\vert{\bf p} - {\bf k}\vert - \lambda)
\nonumber\\
&&\times\bar{\psi}_{\sigma_1}^\alpha (k)\,\bar{\psi}_{\sigma_2}^\alpha (p+q)\,
      \psi_{\sigma_2}^\alpha(k+q)\,\psi_{\sigma_1}^\alpha (p)\quad,
\label{eq:2.2c}
\end{eqnarray}
\begin{eqnarray}
\tilde{S}_{\rm int}^{\,\alpha\,(3)}&=&-\frac T2\sum_{\sigma _1\neq \sigma
_2}\sum_{k,p}{\sum_q}^{\,\prime}
v({\bf k}+{\bf p})\,\Theta (\vert{\bf p} + {\bf k}\vert - \lambda)
\nonumber\\ 
&&\times\bar{\psi}_{\sigma_1}^\alpha (-k)\,\bar{\psi}_{\sigma_2}^\alpha (k+q)\,
     \psi_{\sigma_2}^\alpha (-p+q)\,\psi_{\sigma_1}^\alpha (p)\quad.
\nonumber\\
\label{eq:2.2d}
\end{eqnarray}
\end{mathletters}%
As in real space, we use a $(d+1)$-vector notation with
$k=({\bf k},\omega_n)$, where $\omega_n = 2\pi T(n+1/2)$ is a fermionic
Matsubara frequency.
The prime on the $q$-summation indicates that only momenta up
to the momentum cutoff $\lambda$ are integrated over. While the
long-wavelength, small-frequency phenomena we are interested in do
in general not depend on $\lambda$, the choice of this cutoff will be
important for the range of validity of the final effective theory.
We will come back to this point in Secs.\ \ref{sec:III} and \ref{sec:V}
below. Equations (\ref{eqs:2.2}) represent the same
phase space decomposition as in Eqs.\ (2.8) of I, except that we have
explicitly inserted the step functions in Eqs.\ (\ref{eq:2.2c},\ref{eq:2.2d}),
because the small-wavenumber part of $v$ is already contained in
Eq.\ (\ref{eq:2.2b}). In I the overcounting that resulted from not having
the step functions explicitly present was of no consequence, since $v$
in that case was not singular in the small-wavenumber limit. For a Coulomb
interaction, more care must be taken.  

We next introduce spinors
\begin{mathletters}
\label{eqs:2.3}
\begin{equation}
\psi_{n}^{\alpha}({\bf x}) = \left( \begin{array}{c}
           \psi_{n\uparrow }^{\alpha}({\bf x})\\
           \psi_{n\downarrow}^{\alpha}({\bf x})\end{array}\right)\quad,
\label{eq:2.3a}
\end{equation}
and their Fourier transforms
\begin{equation}
\psi^{\alpha}(k)\equiv\psi_{n}^{\alpha}({\bf k}) = \left(\begin{array}{c} 
                       \psi_{n\uparrow}^{\alpha}({\bf k})\\
                       \psi_{n\downarrow}^{\alpha}({\bf k})\end{array}\right)
                                                                         \quad,
\label{eq:2.3b}
\end{equation}
\end{mathletters}
as well as their adjoints, $\bar\psi^{\alpha}(k)$,
and a scalar product in spinor space, $(\psi ,\psi )=\bar{\psi}\cdot\psi$, 
where the dot denotes the matrix product. Then we can
write the interaction term as
\begin{mathletters}
\label{eqs:2.4}
\begin{equation}
\tilde{S}_{\rm int}^{\,\alpha} = \tilde{S}_{\rm int}^{\,\alpha\,(s)} + 
   \tilde{S}_{\rm int}^{\,\alpha\,(t)} + \tilde{S}_{\rm int}^{\,\alpha\,(3)}
                                                                       \quad, 
\label{eq:2.4a}
\end{equation}
with,
\begin{eqnarray}
\tilde{S}_{\rm int}^{\,\alpha\,(s)}&=&-\frac T2\sum_{k,p}{\sum_q}^{\,\prime}
\Gamma_{k,p}^{(s)}(q)\ \bigl(\psi ^\alpha (k),s_0\psi ^\alpha (k+q)\bigr)
\nonumber\\
&&\times \bigl(\psi ^\alpha (p+q),s_0\psi ^\alpha (p)\bigr)\quad, 
\label{eq:2.4b}
\end{eqnarray}
\begin{eqnarray}
\tilde{S}_{\rm int}^{\,\alpha\,(t)}&=&-\frac T2\sum_{k,p}{\sum_q}^{\,\prime}
\Gamma _{k,p}^{(t)}(q)\sum_{i=1}^3 
\bigl(\psi ^\alpha (k),s_i\psi ^\alpha (k+q)\bigr)
\nonumber\\
&&\times \bigl(\psi ^\alpha (p+q),s_i\psi ^\alpha (p)\bigr)\quad. 
\label{eq:2.4c}
\end{eqnarray}
\end{mathletters}%
Here $s_j=i\sigma_j$, with $\sigma_{1,2,3}$ the Pauli matrices, and
$s_0=\sigma_0$ is the $2\times 2$ identity matrix. We have also defined
the singlet ($s$) and triplet ($t$) interaction amplitudes
\begin{mathletters}
\label{eqs:2.5}
\begin{equation}
\Gamma _{k,p}^{(t)}(q)=\frac{1}{2}\,v({\bf p}-{\bf k})\,
                           \Theta (\vert{\bf p} - {\bf k}\vert - \lambda)\quad,
\label{eq:2.5a}
\end{equation}
and
\begin{equation}
\Gamma_{k,p}^{(s)}(q) = v({\bf q})-\Gamma_{k,p}^{(t)}(q) \quad.
\label{eq:2.5b}
\end{equation}
In addition we define the Cooper channel or $2k_F$-scattering amplitude
\begin{equation}
\Gamma_{k,p}^{(c)}(q) = v({\bf k+p})\,
                        \Theta (\vert{\bf p} + {\bf k}\vert - \lambda)\quad.
\label{eq:2.5c}
\end{equation}
\end{mathletters}%
These expressions are the same as the corresponding ones in I, with the
exception of the momentum restrictions in the effective interaction
potentials $\Gamma_{k,p}^{(s,t,c)}$ discussed above.
Next we project the modes in 
$S_{\rm int}$ onto density modes in the particle-hole and particle-particle
channels that were defined in I. The result of this
procedure, which was explained in Appendix A of I, is
\begin{mathletters}
\label{eqs:2.6}
\begin{eqnarray}
\tilde{S}_{\rm int}^{\,\alpha\,(s)}&=&-\frac T2\sum_{k,p}{\sum_q}^{\,\prime}
\Gamma^{(s)}({\bf q})\ \bigl(\psi ^\alpha (k),s_0\psi ^\alpha (k+q)\bigr)
\nonumber\\
&&\times \bigl(\psi ^\alpha (p+q),s_0\psi ^\alpha (p)\bigr)\quad,
\label{eq:2.6a}
\end{eqnarray}
\begin{eqnarray}
\tilde{S}_{\rm int}^{\,\alpha\,(t)}&=&-\frac{T}{2}\,\Gamma^{(t)}
       \sum_{k,p}{\sum_q}^{\,\prime}\sum_{i=1}^3
        \bigl(\psi^\alpha (k),s_i\psi ^\alpha (k+q)\bigr)
\nonumber\\
&&\times \bigl(\psi ^\alpha (p+q),s_i\psi ^\alpha (p)\bigr)\quad.
\label{eq:2.6b}
\end{eqnarray}
\begin{eqnarray}
\tilde{S}_{int}^{\,\alpha\,(3)}&=&-\frac{T}{2}\,\Gamma^{(c)}
   \sum_{\sigma_1\neq \sigma_2}\sum_{k,p}{\sum_q}^{\,\prime}
     \,\bar{\psi}_{\sigma _1}^\alpha (k)\,\bar{\psi}_{\sigma_2}^\alpha (-k+q)\,
\nonumber\\
&&\times\psi _{\sigma _2}^\alpha (p+q)\,\psi _{\sigma _1}^\alpha (-p)\quad.
\label{eq:2.6c}
\end{eqnarray}
Here 
\begin{equation}
\Gamma^{(s)}({\bf q}) = v({\bf q}) - \Gamma^{(t)}\quad,
\label{eq:2.6d}
\end{equation}
\end{mathletters}%
and $\Gamma^{(t)}$ and $\Gamma^{(c)}$ are numbers that result from
integrating over the wavevectors in Eqs. (\ref{eqs:2.5}) as explained
in Appendix A of I. Notice that as a result of this procedure,
$\Gamma^{(t)}$ and $\Gamma^{(c)}$ for clean electrons depend 
logarithmically on the cutoff $\lambda$, and
diverge as $\lambda\rightarrow 0$. For the disordered case, the logarithmic
singularity is protected both by $\lambda$ and by the disorder. This
singularity, which is a consequence of the Coulomb interaction, is the
reason for our modification of the procedure employed in I. 

\subsection{Composite variables: Matrix field theory}
\label{subsec:II.B}

After completing the phase space decomposition and projecting onto 
densities we are in a position to reformulate the theory in terms 
of composite variables. This proceeds in exact analogy to I, and
we therefore only quote the result. The partition function $\tilde Z$ for the
replicated theory is written as an integral with respect to two
matrix fields, $Q$ and $\tilde\Lambda$,
\begin{equation}
\tilde{Z} = \int D[Q]\,D[\tilde\Lambda]\ e^{{\cal A}[Q,\tilde\Lambda]}
   \quad,
\label{eq:2.7}
\end{equation}
with an effective action
\begin{eqnarray}
{\cal A}[Q,\tilde\Lambda] &=&{\cal A}_{dis}[Q] + {\cal A}_{int}[Q]+\frac{1}{2}\,
   \Tr\ln\left(G_0^{-1}-i\tilde\Lambda\right)
\nonumber\\
&&+\int d{\bf x}\ \tr\left(\tilde\Lambda({\bf x})\,Q({\bf x})\right)\quad.
\label{eq:2.8}
\end{eqnarray}
The matrix field $Q$ corresponds to expressions that are bilinear in the
fermionic fields $\psi$ and $\bar\psi$. Correspondingly, $Q$ carries two
frequency indices $n$ and $m$, and two replica indices $\alpha$ and $\beta$.
Each matrix element $Q_{nm}^{\alpha\beta}$ is an element of ${\cal Q}\times
{\cal Q}$, with $\cal Q$ the quaternion field. $\tilde\Lambda$ is an
auxiliary field whose technical role is to constrain the products of fermionic
fields to the $Q$. It is convenient to expand the $Q_{nm}^{\alpha\beta}$ in 
a spin-quaternion basis,
\begin{equation}
Q_{nm}^{\alpha\beta} = \sum_{r,i=0}^{4}{^i_rQ}_{nm}^{\alpha\beta}\,\tau_r
                       \otimes s_i\quad,
\label{eq:2.8'}
\end{equation}
and analogously for $\tilde\Lambda$. Here $\tau_0 = s_0 = \openone_2$ with
$\openone_2$ the $2\times 2$ 
unit matrix, and $\tau_j = -s_j = -i\sigma_j\ (j=1,2,3)$, where the 
$\sigma_j$ are the Pauli matrices.
The properties of $Q$ and $\tilde\Lambda$ have been derived in I, and for
completeness we list them again in Appendix \ref{app:A}.
In Eq.\ (\ref{eq:2.8}) and in what follows, 
$\Tr$ is a trace over all degrees of freedom, including an integral over 
${\bf x}$, while $\tr$ is a trace over all
discrete degrees of freedom that are not shown explicitly.
\begin{equation}
G_0^{\,-1} = -\partial_{\tau} +\nabla^2 /2m+\mu \quad,
\label{eq:2.9}
\end{equation}
is the inverse of the free electron Green operator, and it is clear
from the structure of the $\Tr\ln$-term in Eq.\ (\ref{eq:2.8}) that the
physical interpretation of the field $\tilde\Lambda$ is that of a self-energy. 
The contributions ${\cal A}_{\rm dis}$ and ${\cal A}_{\rm int}$ 
to the action read
\begin{mathletters}
\label{eqs:2.10}
\begin{equation}
{\cal A}_{\rm dis}[Q] = {\cal A}_{\rm dis}^{\,(1)}[Q] 
                                   + {\cal A}_{\rm dis}^{\,(2)}[Q]\quad, 
\label{eq:2.10a}
\end{equation}
\begin{equation}
{\cal A}_{\rm dis}^{\,(1)}[Q] = \frac{-1}{2\pi N_F\tau_{1}}\int d{\bf x}\ 
                       \Bigl(\tr Q({\bf x})\Bigr)^2\quad, 
\label{eq:2.10b}
\end{equation}
\begin{equation}
{\cal A}_{\rm dis}^{\,(2)}[Q] = \frac{1}{\pi N_F\tau_{\rm rel}}\int d{\bf x}\ 
                       \tr \bigl(Q({\bf x})\bigr)^2\quad, 
\label{eq:2.10c}
\end{equation}
\end{mathletters}%
with $\tau_{\rm rel}$ the single-particle relaxation time and $\tau_1$ a
related scattering time defined in I, and
\begin{mathletters}
\label{eqs:2.11}
\begin{equation}
{\cal A}_{\rm int}[Q] = {\cal A}_{\rm int}^{\,(s)} + {\cal A}_{\rm int}^{\,(t)} 
                                           + {\cal A}_{\rm int}^{\,(c)}\quad, 
\label{eq:2.11a}
\end{equation}
\begin{eqnarray}
{\cal A}_{\rm int}^{\,(s)}&=&\frac{T}{2}\int d{\bf x}d{\bf y}\sum_{r=0,3}(-1)^r
\sum_{n_1,n_2,m}\sum_\alpha  \Gamma^{(s)}({\bf x}-{\bf y})
\nonumber\\
&&\times\left[\tr \left((\tau_r\otimes s_0)\,Q_{n_1,n_1+m}^{\alpha\alpha}
({\bf x})\right)\right] 
\nonumber\\
&&\times\left[\tr \left((\tau_r\otimes s_0)\,Q_{n_2+m,n_2}^{\alpha\alpha}
({\bf y})\right)\right]\quad,
\label{eq:2.11b}
\end{eqnarray}

\begin{eqnarray}
{\cal A}_{\rm int}^{\,(t)}&=&\frac{T\Gamma^{(t)}}{2}\int d{\bf x}
      \sum_{r=0,3}(-1)^r \sum_{n_1,n_2,m}\sum_\alpha\sum_{i=1}^3
\nonumber\\
&&\times\left[\tr\left((\tau_r\otimes s_i)\,Q_{n_1,n_1+m}^{\alpha\alpha}
({\bf x})\right)\right]
\nonumber\\
&&\times\left[\tr\left((\tau_r\otimes s_i)\,Q_{n_2+m,n_2}^{\alpha\alpha}
({\bf x})\right)\right]\quad,
\label{eq:2.11c}
\end{eqnarray}

\begin{eqnarray}
{\cal A}_{\rm int}^{\,(c)} &=&\frac{T\Gamma^{(c)}}{2}\int d{\bf x}\sum_{r=1,2}
                    \sum_{n_1,n_2,m}\sum_{\alpha}
\nonumber\\
&&\times\left[\tr\left((\tau_r\otimes s_0)\,Q_{n_1,-n_1+m}^{\alpha\alpha}
({\bf x})\right)\right]
\nonumber\\
&&\times\left[\tr\left((\tau_r\otimes s_0)\,Q_{-n_2,n_2+m}^{\alpha\alpha}
({\bf x})\right)\right]\quad.
\label{eq:2.11d}
\end{eqnarray}
\end{mathletters}%
Here $N_F$ is the density of states at the Fermi level in saddle-point
approximation, as defined in Eq.\ (\ref{eq:2.12c}) below.
We have written the action in real space, but one should remember
that all of the fields are restricted to Fourier components with
wavenumbers $\vert{\bf k}\vert < \lambda$.

\subsection{Saddle-point solution, Gaussian approximation, and physical
            correlation functions}
\label{subsec:II.C}

It is easy to see that the Fermi-liquid saddle-point discussed in I
remains a valid saddle-point in the presence of a long-range interaction,
with $\Gamma^{(s)}$ in I replaced by $\Gamma^{(s)}({\bf q}=0)$.
The single-particle Green function in saddle-point approximation is
therefore given by the same expression as in I, viz.
\begin{mathletters}
\label{eqs:2.12}
\begin{equation}
G_{\rm sp}({\bf p},\omega_n) = \left[i\omega_n - {\bf p}^2/2m + \mu -
   \Sigma_n\right]^{-1}\quad,
\label{eq:2.12a}
\end{equation}
with $\mu$ the chemical potential, and
the self-energy $\Sigma$ the solution of the equation
\begin{eqnarray}
\Sigma_n&=&\frac{1}{\pi N_F\tau_{\rm rel}}\,\frac{1}{V}\,
      \sum_{\bf p}\,\left[ i\omega_n
    - {\bf p}^2/2m + \mu - \Sigma_{n}\right]^{-1}
\nonumber\\
        &&+2\Gamma^{(s)}({\bf q}=0) T\sum_{m}\,e^{i\omega_m 0}\,\frac{1}{V}\,
          \sum_{\bf p}\,\bigl[i\omega_{m}
          - {\bf p}^2/2m
\nonumber\\
        &&\qquad\qquad\qquad\qquad\qquad + \mu - \Sigma_{m}\bigr]^{-1}\quad.
\label{eq:2.12b}
\end{eqnarray}
This is the standard Hartree-Fock result, with the disorder treated in
self-consistent Born approximation.
\begin{equation}
N_F = \frac{-2}{\pi}\,\frac{1}{V}\,\sum_{\bf p} {\rm Im}\,
    G_{\rm sp}({\bf p},i\omega_n\rightarrow i0)\quad,
\label{eq:2.12c}
\end{equation}
\end{mathletters}%
is the density of states at the Fermi level in saddle-point approximation,
which is used for normalization purposes throughout.

Similarly, in an expansion about the saddle-point to Gaussian order, all
of the expressions derived in I remain valid if we substitute 
$\Gamma^{(s)}({\bf p})$ for $\Gamma^{(s)}$ in all propagators taken at
wavevector ${\bf p}$. In particular we obtain for the low-frequency,
long-wavelength limit of the density susceptibility, $\chi$, in the
disordered case
\begin{mathletters}
\label{eqs:2.13}
\begin{equation}
\chi({\bf p},\Omega_n) = \chi_{\rm st}({\bf p})\,\frac{d({\bf p})
                                                                 {\bf p}^2}
                           {\vert\Omega_n\vert + d({\bf p}){\bf p}^2} \quad.
\label{eq:2.13a}
\end{equation}
Here $\Omega_n$ is a bosonic Matsubara frequency.
\begin{equation}
\chi_{\rm st}({\bf p}) = \frac{-N_F}{1 + N_F\Gamma^{(s)}({\bf p})}\quad,
\label{eq:2.13b}
\end{equation}
is the static density susceptibility, and 
\begin{equation}
d({\bf p}) = D[1+N_F\Gamma^{(s)}({\bf p})]\quad,
\label{eq:2.13c}
\end{equation}
\end{mathletters}%
with $D$ the Boltzmann diffusion coefficient. In the clean limit we obtain
the usual RPA expression,
\begin{equation}
\chi({\bf p},\Omega_n) = \frac{\chi_0({\bf p},\Omega_n)}
   {1 + \Gamma^{(s)}({\bf p})\chi_0({\bf p},\Omega_n)}\quad,
\label{eq:2.14}
\end{equation}
with $\chi_0$ the Lindhard function.
We see that in Gaussian approximation, the field theory describes screening,
and the existence of plasmons, in RPA and its disordered generalization,
respectively.

\section{Symmetry analysis}
\label{sec:III}

From Sec.\ \ref{subsec:II.C} it follows, in conjunction with I, that at
the level of the Gaussian approximation the $Q_{nm}$ 
with $nm<0$ are soft modes. In this section, we perform
a symmetry analysis in order to show that they are indeed the exact
soft modes of the theory. This
will allow us to explicitly separate the soft modes from the massive ones,
and to formulate an effective field theory for which the soft modes remain
manifestly soft to all orders in perturbation theory.

\subsection{Basic transformation properties, and Ward identity}
\label{subsec:III.A}

Let us start with a symmetry analysis of our field theory that is a slight
generalization of the procedure followed in I, which in turn was a
generalization of the work on noninteracting electrons by Sch{\"a}fer
and Wegner, and Pruisken and Sch{\"a}fer.\cite{SchaferWegner} We consider 
an infinitesimal simultaneous rotation in frequency and replica space
space given by,
\begin{eqnarray}
^i_r{\hat T}_{nm}^{\alpha\beta}&=&\delta_{i0}\,\delta_{r0}\,
               \Bigl[     \delta_{\alpha\alpha_1}\delta_{n n_1}
                          \delta_{\beta\alpha_2}\delta_{m n_2}
\nonumber\\
           &&\qquad\qquad - \delta_{\alpha\alpha_2}\delta_{n n_2}
                          \delta_{\beta\alpha_1}\delta_{m n_1}
\Bigr]\theta
\nonumber\\
  &\equiv&\delta_{i0}\,\delta_{r0}\,{\hat t}_{nm}^{\alpha\beta}
               + O(\theta^2)\quad,
\label{eq:3.1}
\end{eqnarray}
which transforms the $Q$-matrices according to $Q\rightarrow TQT^{-1}$,
with $T=C{\hat T}C^T$, where $C=i\tau_1\otimes s_2$. $(n_1>0,n_2<0)$ 
and $(\alpha_1,\alpha_2)$ are fixed pairs of frequency and replica indices
that characterize the transformation. Under such an
infinitesimal rotation, the $Q$-matrices transform like
\begin{mathletters}
\label{eqs:3.2}
\begin{equation}
Q_{nm}^{\alpha\beta} \rightarrow Q_{nm}^{\alpha\beta}
                   + \delta Q_{nm}^{\alpha\beta}\quad,
\label{eq:3.2a}
\end{equation}
with
\begin{eqnarray}
\delta Q_{nm}^{\alpha\beta} = \bigl(\delta_{\alpha\alpha_1}\delta_{n n_1}
Q_{n_2 m}^{\alpha_2\beta}
+ \delta_{\beta\alpha_1}\delta_{m n_1} Q_{n n_2}^{\alpha\alpha_2}
\nonumber\\
- (1 \leftrightarrow 2) \bigr)\ \theta
+ O(\theta^2)\quad.
\label{eq:3.2b}
\end{eqnarray}
\end{mathletters}%
Here we have shown only the frequency and replica indices, since all 
other degrees of
freedom are unaffected by the transformation. The $\tilde\Lambda$-matrices
transform accordingly. The symbol $(1\leftrightarrow 2)$ indicates the same
terms as written previously, but with the indices 1 and 2 interchanged.

Of the terms in the action, Eq.\ (\ref{eq:2.8}), only 
$\Tr \ln (G_0^{-1} - i\tilde\Lambda)$ and
${\cal A}_{int}$ are not invariant under the above transformation. Their
transformation behaviors are easily determined by an explicit calculation.
We find
\begin{mathletters}
\label{eqs:3.3}
\begin{equation}
\Tr \ln \left(G_0^{-1} - i\tilde\Lambda\right) \rightarrow
   \Tr \ln \left(G_0^{-1} - i\tilde\Lambda\right)
        + \theta\,\Tr (G\,\delta i\omega)\quad,
\label{eq:3.3a}
\end{equation}
with $G \equiv (G_0^{-1} - i\tilde\Lambda)^{-1}$, and
\begin{eqnarray}
^i_r(\delta i\omega)_{nm}^{\alpha\beta} &=& \delta_{i0}\,\delta_{r0}\,
   \left(\delta_{\alpha\alpha_1}\delta_{n n_1}
         \delta_{\beta\alpha_2}\delta_{m n_2} +
         \delta_{\alpha\alpha_2}\delta_{n n_2} \right.
\nonumber\\
         &&\left.\times\delta_{\beta\alpha_1} \delta_{m n_1}\right)\,
                       i\Omega_{n_1 - n_2}\quad,
\label{eq:3.3b}
\end{eqnarray}
\end{mathletters}%
and
\begin{mathletters}
\label{eqs:3.4}
\begin{equation}
{\cal A}_{int}^{\,(s)}\rightarrow {\cal A}_{int}^{\,(s)}
+ \delta {\cal A}_{int}^{\,(s)}\quad,
\label{eq:3.4a}
\end{equation}
with
\begin{eqnarray}
\delta {\cal A}_{int}^{\,(s)} &=& 32 \int d{\bf x} d{\bf y}\,\Gamma^{(s)}
                 ({\bf x- y}) \sum_{r=0,3}
\nonumber\\
&&\times T\sum_{{n}_a {n}_b}\Bigl[
               {^{0}_{r}Q}_{{n}_a {n}_b}^{\alpha_1\alpha_1}({\bf x})\,
               {^{0}_{r}Q}_{n_2,n_2-({n}_a-{n}_b)}^{\alpha_2\alpha_1}
                 ({\bf y}) 
\nonumber\\
&&\qquad\qquad - (1\leftrightarrow 2) \Bigr]\,\theta\quad.
\label{eq:3.4b}
\end{eqnarray}
\end{mathletters}%
For our purposes, we concentrate on a discussion of the particle-hole
spin-singlet interaction; the discussion of the remaining interaction
channels proceeds analogously, but is less interesting since in these
channels the interaction is short-ranged. Proceeding analogously to I, 
we obtain from the above transformation properties the Ward identity
\begin{mathletters}
\label{eqs:3.5}
\begin{eqnarray}
W_{int}&+&8\Omega_{n_1 - n_2}\int d{\bf y}\
      \left\langle{^{0}_{0}Q}_{1 2}({\bf y})\
              ^{0}_{0}Q_{3 4}({\bf x})\right\rangle
\nonumber\\
&=& \delta_{1 3}\delta_{2 4}\left(\left\langle{^{0}_{0}Q}_{1 1}
  ({\bf x})\right\rangle - \left\langle{^{0}_{0}Q}_{2 2}
                 ({\bf x})\right\rangle\right)\quad,
\nonumber\\
\label{eq:3.5a}
\end{eqnarray}
where,
\begin{eqnarray}
W_{int}&=&
-32\int d{\bf y} d{\bf z}\ \Gamma^{(s)}({\bf y} -{\bf z})
\sum_{r=0,3}\ T\sum_{{n}_a {n}_b}
\nonumber\\
&&   \biggl[\Bigl\langle{^{0}_{0}Q}_{3 4}({\bf x})
{^{0}_{r}Q}_{{n}_a {n}_b}^{\alpha_1\alpha_1}({\bf y})\,
   {^{0}_{r}Q}_{n_2,n_1-({n}_a-{n}_b)}^{\alpha_2\alpha_1}({\bf z})
    \Bigr\rangle
\nonumber\\
&& - (1\leftrightarrow 2) \biggr]\quad,
\label{eq:3.5b}
\end{eqnarray}
\end{mathletters}%
To simplify the notation we have adopted the
convention $1\equiv(n_1,\alpha_1)$. For the special case of a 
short-ranged interaction,
$\Gamma^{(s)}({\bf x}-{\bf y}) = \Gamma^{(s)}\,\delta({\bf x}-{\bf y})$, 
we recover Eqs.\ (3.14) of I.

\subsection{Solution of the Ward identity}
\label{subsec:III.B}

We now solve the Ward identity by employing a method that is more transparent
than the one used in I, and more suitable for a generalization to the
long-range case.

We first split up the $Q$-matrices into their averages and fluctuations:
$Q_{nm} = \langle Q_{nm}\rangle + (\Delta Q)_{nm}
= \delta_{nm}\langle Q_{nn}\rangle + (\Delta Q)_{nm}$.
Since $n_1\neq n_2$, and
$n_3\neq n_4$, we have
$\langle Q_{n_1 n_2}\rangle = \langle Q_{n_3 n_4}\rangle = 0$. Furthermore,
if we put $n_a = n_b$ in the 3-point functions, then the expression in
square brackets in Eq.\ (\ref{eq:3.5b}) vanishes
due to Eq.\ (\ref{eq:A.3a}), so effectively we also
have $\langle Q_{n_a n_b}\rangle = 0$. Equation\ (\ref{eq:3.5a}) then
takes the form,
\begin{mathletters}
\label{eqs:3.6}
\begin{equation}
W_{int}+8\Omega_{n_1 - n_2} C_{1 2, 3 4} = \delta_{13}\,\delta_{24}\,N_{12}
               \quad,
\label{eq:3.6a}
\end{equation}
where
\begin{equation}
C_{12,34} = \int d{\bf y}\
                \biggl\langle{^{0}_{0}(\Delta Q)}_{1 2}({\bf y})\,
                {^{0}_{0}(\Delta Q)}_{3 4}({\bf x})\biggr\rangle\quad,
\label{eq:3.6b}
\end{equation}
and
\begin{equation}
N_{1 2} = \Bigl\langle{^{0}_{0}Q}_{1 1}({\bf x})\Bigr\rangle
        - \Bigl\langle{^{0}_{0}Q}_{2 2}({\bf x})\Bigr\rangle
\quad.
\label{eq:3.6c}
\end{equation}
$W_{\rm int}$ can be written as the sum of two terms,
$W_{\rm int} = W_{\rm int}^{(1)} + W_{\rm int}^{(2)}$, with
\begin{eqnarray}
W_{\rm int}^{(1)}&=&-32 \delta_{\alpha_1\alpha_2}
\Gamma^{(s)}({\bf k}\rightarrow 0) N_{12}
\nonumber\\
&&\times T\sum_{n_a,n_b}\delta_{\alpha_1\alpha_a}\delta_{a-b,1-2}\,C_{ab,34}
                                  \quad,
\label{eq:3.6d}
\end{eqnarray}
and
\begin{eqnarray}
W_{\rm int}^{(2)}&=&-32 \sum_{r=0,3}\ {\sum_{\bf q}}^{\,\prime}\
\Gamma^{(s)}({\bf q})\ T\ \sum_{n_a n_b}
\nonumber\\
 &&\biggr[\Bigl<{^{0}_{r}(\Delta Q)}_{n_1,n_2-(n_a-n_b)}
                       ^{\alpha_1\alpha_2}({\bf -q})\
                 {^{0}_{r}(\Delta Q)}_{n_a n_b}^{\alpha_1\alpha_1}({\bf q})
\nonumber\\
&&\qquad\times {^{0}_{0}(\Delta Q)}_{3 4}({\bf x})\Bigr> 
       - (1\leftrightarrow 2)\biggr] \quad,     
\label{eq:3.6e}
\end{eqnarray}
\end{mathletters}%
Here we have chosen a mixed representation for
$W_{\rm int}^{(2)}$, with some $\Delta Q$ in real space, and some in Fourier
space, in order to make the cutoff on the ${\bf q}$-integration explicit.

To proceed, let us ignore $W_{\rm int}^{(2)}$ for the time being. Its effect
will be analyzed later. Equation (\ref{eq:3.6a}) then turns into a closed
integral equation for the homogeneous correlation function $C$. At this
point we note that our global symmetry transformation, Eq.\ (\ref{eq:3.1}),
produces a Ward identity for homogeneous correlation functions. For a
short-ranged interaction, this is sufficient to capture the important
structural restrictions imposed on the theory by the symmetry of the action.
However, in the long-range case the homogeneous limit is singular, and we
have to be more careful. For instance, it is obvious that a local
symmetry transformation would generate a wavenumber dependent
$\Gamma^{(s)}$ in $W_{\rm int}^{(1)}$, and in the long-range case
$\Gamma^{(s)}({\bf k}\rightarrow 0) \neq \Gamma^{(s)}({\bf k} = 0)$.
This is the reason why we have written $\Gamma^{(s)}({\bf k}\rightarrow 0)$
in Eq.\ (\ref{eq:3.6d}). Furthermore, it is known from perturbation theory
that the dispersion of the soft modes that are controlled by the Ward
identity is diffusive, $\Omega\sim {\bf k}^2$.\cite{DiffusionFootnote} The
equation for the wavenumber dependent $C$ then takes the form
\begin{eqnarray}
8(\Omega_{n_1-n_2}&+&D{\bf k}^2)\,C_{12,34}({\bf k}) = 
   \delta_{13}\,\delta_{24}\,N_{12}({\bf k})
\nonumber\\
&+&32\Gamma^{(s)}({\bf k})\,\delta_{\alpha_1\alpha_2}\,N_{12}({\bf k})\,
\nonumber\\
&&\times T\sum_{ab}\delta_{\alpha_a\alpha_1}\delta_{1-2,a-b}\, C_{ab,34}({\bf k})
                         \quad,
\label{eq:3.7}
\end{eqnarray}
with $D$ the (exact) diffusion constant, and $N_{12}({\bf k})$ the
wavevector dependent generalization of $N_{12}$ as defined in 
Eq.\ (\ref{eq:3.6c}). It is useful to define
\begin{mathletters}
\label{eqs:3.8}
\begin{equation}
P_n({\bf k}) = T\sum_{n_1 n_2} \delta_{n,n_1-n_2}\,N_{n_1n_2}({\bf k})\quad.
\label{eq:3.8a}
\end{equation}
By performing appropriate summations over the Ward identity, we find that
\begin{equation}
P_n({\bf k}) = \frac{1}{4}\,\left(\Omega_n + D{\bf k}^2\right)\,
      \chi_{\rm sc}({\bf k},\Omega_n)\quad,
\label{eq:3.8b}
\end{equation}
with $\chi_{\rm sc}$ the screened density susceptibility. The latter is
defined as
\begin{equation}
\chi_{\rm sc}({\bf k},\Omega_n) = \frac{\chi({\bf k},\Omega_n)}
  {1 + \Gamma^{(s)}({\bf k})\chi({\bf k},\Omega_n)}\quad,
\label{eq:3.8c}
\end{equation}
where
\begin{equation}
\chi({\bf k},\Omega_n) = 32T\sum_{1234} \delta_{n,n_1-n_2}\,C_{12,34}({\bf k}) 
       \quad,
\label{eq:3.8d}
\end{equation}
\end{mathletters}%
is the full density susceptibility.
Notice that this relation between $P_n$ and $\chi_{\rm sc}$
is exact, since $T\sum_{12}\delta_{n,n_1-n_2} W^{(2)}_{12,34} = 0$.

The integral equation, Eq. (3.7), can be solved in terms of $P_n$
or $\chi_{\rm sc}$ by means of the same methods that were
employed to discuss the Gaussian $Q$-field theory in I. We find
\begin{mathletters}
\label{eqs:3.9}
\begin{eqnarray}
C_{12,34}({\bf k})&=&\frac{1}{16}\,\Bigl[\delta_{13}\delta_{24}\,{\cal D}_{12}
  ({\bf k})
\nonumber\\
&& + \delta_{1-2,3-4}\delta_{\alpha_1\alpha_3}\delta_{\alpha_1\alpha_2}
               \,2T\Gamma^{(s)}({\bf k})
\nonumber\\
&&\qquad\times {\cal D}_{12}({\bf k})\,{\cal D}^{(s)}_{34}({\bf k})\Bigr]\quad,
\label{eq:3.9a}
\end{eqnarray}
where
\begin{equation}
{\cal D}_{12}({\bf k}) = 2N_{12}({\bf k})\,\left[\Omega_{n_1-n_2} + D{\bf k}^2
                                    \right]^{-1}\quad,
\label{eq:3.9b}
\end{equation}
and
\begin{equation}
{\cal D}^{(s)}_{12}({\bf k}) = {\cal D}_{12}({\bf k})\,\left[1 -
    \Gamma^{(s)}({\bf k})\chi_{\rm sc}({\bf k},\Omega_{n_1-n_2})\right]^{-1}\ .
\label{eq:3.9c}
\end{equation}
\end{mathletters}%
To appreciate the difference between this structure and the analogous one
in the short-range case, it is instructive to consider the limit of small
wavenumbers and frequencies with $D{\bf k}^2 \ll \Omega_{n_1-n_2}$. In
this limit we have $2N_{12}({\bf k})\rightarrow \pi\,N(\epsilon_F)$ with
$N(\epsilon_F)$ the exact density of states at the Fermi level, and
$\chi_{\rm sc}({\bf k},\Omega_{n_1-n_2})\sim {\bf k}^2/\Omega_{n_1-n_2}$.
In $d=3$, we obtain
\begin{eqnarray}
16\,C_{12,34}({\bf k}&\rightarrow&0) = \delta_{12}\delta_{34}\,
         \frac{\pi N(\epsilon_F)}{\Omega_{n_1-n_2}}
\nonumber\\
&+& \delta_{1-2,3-4}\,\delta_{\alpha_1\alpha_2}\,\delta_{\alpha_1\alpha_3}\,
      \frac{\pi N(\epsilon_F)}{\Omega_{n_1-n_2}}\,
      \frac{2\pi T}{D{\bf k}^2}\ .
\label{eq:3.10}
\end{eqnarray}
We see that the long-ranged Coulomb interaction causes the 
part of $C$ that is non-diagonal in
frequency to diverge like $T/\Omega\,{\bf k}^2$, rather than
like $T/\Omega^2$ in the short-range case. This structure is responsible
for the well-known log-squared terms in the density of states and in the
sound attenuation coefficient that appear in perturbative calculations of
weak-localization effects in $2$-$d$ systems with Coulomb interactions.\cite{R}

We now consider the remaining contribution to $W_{\rm int}$, 
$W_{\rm int}^{(2)}$.
Without the restriction on the wavenumber integral in Eq.\ (\ref{eq:3.6e}),
one could relabel wavevectors to show that $W_{\rm int}^{(2)}$ contains terms
that have the same structure as $W_{\rm int}^{(1)}$. However, due to the
cutoff contained in the definition of our $Q$-field theory this is not
the case, and we resort to perturbation theory to analyze the structure
of $W_{\rm int}^{(2)}$. By re-expressing $\Delta Q$ in fermion fields, and
using Wick's theorem to write the correlation function in Eq.\ (\ref{eq:3.6e})
in terms of Green functions, we find
\begin{equation}
W_{\rm int}^{(2)} \sim \Gamma^{(s)}({\bf k})\,\frac{T}{\Omega}\,{\rm Min}
       \left(\lambda^2,\lambda^2 (\lambda\ell)\right)\quad,
\label{eq:3.11}
\end{equation}
with $\lambda$ the cutoff introduced in Sec.\ \ref{subsec:II.A}, and
$\ell$ the elastic scattering mean-free path. Based on this information,
we now choose the cutoff. There are three obvious possible choices: One
can make $\lambda$ a fixed, small fraction of either the Fermi wavenumber
$k_F$, or the mean-free path $\ell$, or the screening wavenumber $\kappa$.
The first choice would not allow us to control the terms in
$W_{\rm int}^{(2)}$, and is therefore undesirable. The second choice is
possible, but means foregoing the option to take the clean limit, for
reasons discussed after Eq.\ (\ref{eq:2.6d}). We therefore choose the
third option, which makes
$W_{\rm int}^{(2)}$ of higher order in the interaction than the other terms
in the Ward identity. To linear order in the interaction the
solution given above in Eqs.\ (\ref{eqs:3.9}), (\ref{eq:3.10}), 
is then exact. To higher order,
while $W_{\rm int}^{(2)}$ cannot make the modes $Q_{nm}$ ($nm<0$) massive, it
will change the prefactors of the diffusive singularity at small
frequencies and wavenumbers. We conclude that an effective theory that
respects Eqs.\ (\ref{eqs:3.9}) will have the correct soft-mode structure
as it follows from the symmetry of the action. It will further exactly
reproduce perturbation theory to first order in the interaction.
In Sec.\ \ref{sec:IV} we will show that the generalized nonlinear
$\sigma$ model of Ref.\ \onlinecite{Finkelstein} is an effective theory
with these properties.

\subsection{Separation of soft and massive modes}
\label{subsec:III.D}

From the previous subsection we know that the correlation functions of
the $Q_{nm}$ with $nm<0$ remain
soft, while those with $nm>0$ remain massive even in the presence of 
long-range interactions. Therefore the mode separation for this case
is the same as it is for short-range interactions,
and we can restrict ourselves to summarizing the results of I.

One consequence of the symmetry properties of the $Q$-matrices 
(see Appendix\ \ref{app:A}) is that the set of $Q$ is
isomorphic to the set of antihermitian $8Nn\times 8Nn$ matrices.
As shown in I, from this it follows that the most general $Q$ can
be written as a product,
\begin{equation}
Q = {\cal S}\ P\ {\cal S}^{-1}\quad.
\label{eq:3.12}
\end{equation}
Here $P$ is a matrix that is blockdiagonal in frequency space,
\begin{equation}
P = \left( \begin{array}{cc}
       P^> & 0   \\
       0   & P^< \\
    \end{array} \right)\quad,
\label{eq:3.13}
\end{equation}
with elements $P^{>}$ ($P^{<}$) for $nm>0$ ($nm<0$) that are 
isomorphic to the set of antihermitian $4Nn\times 4Nn$ matrices,
and ${\cal S}$ an element of the homogeneous space
${\rm USp}(8Nn,{\cal C})/{\rm USp}(4Nn,{\cal C})\times {\rm USp}(4Nn,{\cal C})$,
i.e. the set of all cosets of ${\rm USp}(8Nn,{\cal C})$ with respect to
${\rm USp}(4Nn,{\cal C})\times {\rm USp}(4Nn,{\cal C})$.\cite{Gilmore}

This achieves the desired separation of our degrees of freedom into soft
and massive ones. The massive degrees of freedom are represented by the
matrix $P$, while the soft ones are represented by the transformations
${\cal S}\in {\rm USp}(8Nn,{\cal C})/{\rm USp}(4Nn,{\cal C})\times 
{\rm USp}(4Nn,{\cal C})$. 

In order to formulate the field theory in terms of the soft and massive
modes, one also needs the invariant measure $I[P]$, or the Jacobian
of the transformation from the $Q$ to the $P$ and the $\cal S$, defined by
\begin{equation}
\int D[Q]\,\ldots = \int D[P]\,I[P] \int D[\cal S]\,\ldots\quad.
\label{eq:3.24}
\end{equation}
We will not need the measure explicitly for our purposes, and refer to
I, where it has been constructed in terms of the eigenvalues of $P$.

\section{Effective field theory for disordered interacting fermions}
\label{sec:IV}

Having achieved a separation of soft and massive modes, we
are now in a position to formulate an effective theory for electrons
with a long-range interaction that focuses on the soft modes.
In the short-range case, this was done by integrating out the
massive modes in tree approximation. This led to the 
nonlinear $\sigma$ model in I, and it was shown that this procedure
preserves the structure of the Ward identity. As a result, the
$\sigma$ model contains the same Fermi liquid fixed point, as well
as the leading corrections to scaling near it, as the underlying
full model, and it also contains a critical fixed point that
describes an Anderson-Mott metal-insulator transition.

We cannot simply repeat this procedure for the present case of
long-range interactions, since integrating out the massive modes
in tree approximation would lead to a theory that violates the
Ward identity. We therefore must treat the massive modes more
carefully, and our aim is to find the simplest approximation that
will still guarantee the correct structure of the Ward identity,
and hence lead to an effective theory that has the correct symmetry.

The first steps are the same as in I: We define a transformed auxiliary
field $\Lambda$ by
\begin{mathletters}
\label{eqs:4.1}
\begin{equation}
\Lambda ({\bf x}) = {\cal S}^{-1}({\bf x})\,{\tilde\Lambda}({\bf x})\,{\cal S}
                        ({\bf x})\quad,
\label{eq:4.1a}
\end{equation}
and a new field ${\hat Q}$ by
\begin{equation}
{\hat Q}({\bf x}) = \frac{4}{\pi N_F}\,{\cal S}({\bf x})\langle P\rangle 
                    {\cal S}^{-1} ({\bf x})\quad.
\label{eq:4.1b}
\end{equation}
\end{mathletters}%
We then expand about the expectation values of $\Lambda$ and $P$,
\begin{equation}
P = \langle P\rangle + \Delta P\quad,\quad \Lambda = \langle\Lambda\rangle
                                            + \Delta\Lambda\quad.
\label{eq:4.2}
\end{equation}
As explained in I, it is sufficient to replace $\langle P\rangle$ and
$\langle\Lambda\rangle$ by the respective saddle-point values, and
to further replace $\langle P\rangle$ in the definition of ${\hat Q}$,
Eq.\ (\ref{eq:4.1b}), by the simple approximation
\begin{mathletters}
\label{eqs:4.3}
\begin{equation}
\langle P\rangle \approx \frac{\pi}{4}\,N_F\,\pi\quad,
\label{eq:4.3a}
\end{equation}
with
\begin{equation}
\pi_{12} = \delta_{12}\,\sgn\omega_{n_1}\quad.
\label{eq:4.3b}
\end{equation}
\end{mathletters}%
We mention that there is no obvious small parameter that controls these
approximations. Rather, they will be justified a posteriori by the fact
that the resulting effective field theory, the nonlinear $\sigma$ model,
respects the Ward identity, Eqs.\ (\ref{eqs:3.9}), (\ref{eq:3.10}),
that was derived in the previous section. This in turn shows that the
approximations, Eqs.\ (\ref{eqs:4.3}), are consistent with neglecting
the term $W_{\rm int}^{(2)}$ in the Ward identity, which itself is
perturbatively controlled for small interaction strengths. If the
theory is renormalizable, this implies that the effective theory
resulting from the above approximations will have the same structure
as the full one, albeit with different coefficients. We will come
back to this point at the end of this section, and in Sec.\ \ref{sec:V} 
below. With these approximations, ${\hat Q}$ has the properties
\begin{mathletters}
\label{eqs:4.4}
\begin{equation}
{\hat Q}^2({\bf x}) = \openone ,\quad 
                         {\hat Q}^{\dagger} = {\hat Q}\ ,
                            \quad \tr{\hat Q}({\bf x}) = 0\ ,
\label{eq:4.4a}
\end{equation}
and can be parametrized by
\begin{equation}
{\hat Q} = \left( \begin{array}{cc}
                 \sqrt{1-qq^{\dagger}} & q   \\
                    q^{\dagger}        & -\sqrt{1-q^{\dagger} q} \\
           \end{array} \right)\quad,
\label{eq:4.4b}
\end{equation}
where the matrix $q$ has elements $q_{nm}$ whose frequency labels are
restricted to $n\geq 0$, $m<0$. $\cal S$ can also be expressed in term
of $q$,\cite{Gilmore}
\begin{equation}
{\cal S} = \left( \begin{array}{cc}
                 \sqrt{1-bb^{\dagger}} & b   \\
                   -b^{\dagger}        & \sqrt{1-b^{\dagger} b} \\
           \end{array} \right)\quad,
\label{eq:4.4c}
\end{equation}
where
\begin{equation}
b(q,q^{\dagger}) = \frac{-1}{2}\,q\,f(q^{\dagger}q)\quad,
\label{eq:4.4d}
\end{equation}
with
\begin{equation}
f(x) = \sqrt{\frac{2}{x}}\,\left(1 - \sqrt{1 - x}\right)^{1/2}\quad.
\label{eq:4.4e}
\end{equation}
\end{mathletters}%

Unlike I, where we just dropped the fluctuations of $P$ and $\Lambda$, here we
next expand to second order in $\Delta P$ and $\Delta\Lambda$. The reason
for this change of procedure compared to the short-ranged case is that
here the $\Delta P$ fluctuations are multiplied by a divergent Coulomb
potential, and therefore must be retained.
The part of
the action that is quadratic in these massive fluctuations reads
\begin{eqnarray}
{\cal A}_m^{(2)}&=&\frac{1}{4}\int d{\bf x}\,d{\bf y}\,\tr 
  G_{\rm sp}({\bf x}-{\bf y})\,\Delta\Lambda({\bf y})\,
           G_{\rm sp}({\bf y}-{\bf x})\,\Delta\Lambda({\bf x})
\nonumber\\
&& + \int d{\bf x}\,\Delta\Lambda ({\bf x})\,\Delta P({\bf x})
\nonumber\\
&& + {\cal A}_{\rm int}^{(s)}\left[{\cal S}\langle P\rangle{\cal S}^{-1}
             + {\cal S}\Delta P{\cal S}^{-1}\right]\quad,
\label{eq:4.5}
\end{eqnarray}
with ${\cal A}_{\rm int}^{(s)}[Q]$ the spin-singlet interaction term from
Eq.\ (\ref{eq:2.11b}). ${\cal A}_{\rm dis}$, and the invariant measure,
expanded to second order in $\Delta P$, also contribute to 
${\cal A}_m^{(2)}$. However, their net effect is to add a constant to
the singular interaction potential in the ${\cal A}_{\rm int}^{(s)}$
contribution to ${\cal A}_m^{(2)}$, and hence they can be neglected.
Likewise, there are terms linear in $\Delta\Lambda$ that couple to
the soft modes ${\cal S}$. These always contain at least one frequency
or gradient squared, and therefore are unimportant for the leading
structure imposed by the Ward identity. They also turn out to be
renormalization group irrelevant at the Fermi liquid fixed point that
we will discuss in Sec.\ \ref{sec:V} below. We therefore neglect all
of these terms.

Because of the coupling between $\Delta P$ and ${\cal S}$, 
${\cal A}_m^{(2)}$ still represents a complicated quadratic form.
To handle it, we expand $\cal S$, Eq.\ (\ref{eq:4.4c}), in powers of $q$.
To lowest order, we just have ${\cal S} = 1$. It turns out that this
lowest order approximation is sufficient to ensure the correct structure
of the Ward identity. We have also explicitly checked that higher order terms
in this $q$-expansion lead to corrections that are irrelevant near the
Fermi liquid fixed point. With ${\cal S} = 1$, the massive Gaussian
fluctuations are easily integrated out. Neglecting terms that are of
first or higher order in the external frequency, the result is a
change of the interaction term ${\cal A}_{\rm int}^{(s)}$ to a term of
the same structure, but with $\Gamma^{(s)}$ replaced by its screened
counterpart
\begin{equation}
\Gamma_{\rm sc}^{(s)}({\bf p}) = \frac{\Gamma^{(s)}({\bf p})}
             {1 + N_F\Gamma^{(s)}({\bf p})}\quad,
\label{eq:4.6}
\end{equation}
with $N_F$ from Eq.\ (\ref{eq:2.12c}) (here we have neglected a wavenumber
dependence that is subleading compared to that of $\Gamma^{(s)}$). 
We see that integrating out the massive fluctuations
in the approximation we have chosen leads to static screening of the
Coulomb interaction. Analogous screening effects occur in the remaining
interaction channels. However, they are uninteresting there since they
just renormalize the numbers $\Gamma^{(t)}$ and $\Gamma^{(c)}$.

The remaining steps in the derivation of the nonlinear $\sigma$ model
are the same as in I. We thus obtain the $\sigma$ model action
\begin{eqnarray}
{\cal A}_{\rm NL\sigma M}&=&\frac{-1}{2G}\int d{\bf x}\ 
     \tr\left(\nabla\tilde Q ({\bf x})\right)^2
\nonumber\\
&&+ 2H \int d{\bf x}\ \tr\left(\Omega\,{\tilde Q}({\bf x}) \right) 
           + {\cal A}_{\rm int}[{\tilde Q}]\quad,
\label{eq:4.7}
\end{eqnarray}
where ${\tilde Q} = {\hat Q} - \pi$, with $\pi$ from Eq.\ (\ref{eq:4.3b}).
$G = 8/\pi\sigma_0$ with $\sigma_0$ the conductivity in self-consistent
Born approximation, and $H = \pi N_F/8$. ${\cal A}_{\rm int}$ is given
by Eqs.\ (\ref{eqs:2.11}), but with $\Gamma^{(s)}$ replaced by 
$\Gamma_{\rm sc}^{(s)}$, Eq.\ (\ref{eq:4.6}). This is the generalized
nonlinear $\sigma$ model for disordered electrons with a Coulomb interaction,
as proposed and discussed by Finkel'stein,\cite{Finkelstein}
and the above procedure represents a technical derivation
of this model. Its properties have been reviewed
and discussed in detail in Ref.\ \onlinecite{R}. Using the explicit results
of that reference, it is straightforward to check that the model indeed
obeys the Ward identity, Eq.\ (\ref{eq:3.10}).\cite{RenormalizabilityFootnote}
This justifies, a posteriori,
the approximations that have entered our derivation. We will further
discuss the merits and limitations of the model in the next section.

\section{Discussion}
\label{sec:V}

We finally discuss our results, and the procedures used to derive them.

\subsection{Role of the phase space decomposition}
\label{subsec:V.A}

Let us start with a discussion of the phase space decomposition in 
Sec.\ \ref{subsec:II.A}, which writes the interaction term (and also
the disorder term) in the action as a sum of terms of different structures,
with a cutoff to avoid double counting. As a result, the definition of the
effective action contains this cutoff, which is a priori unspecified. 
For instance, without the cutoff $\lambda$ the three terms 
${\tilde S}_{\rm int}^{\alpha\,(1,2,3)}$, Eqs.\ (\ref{eqs:2.2}), would all
be equal and equal to ${\tilde S}_{\rm int}^{\alpha}$. A superficial
consideration might conclude that this decomposition of the action
introduces an unnecessary ambiguity into the theory. 
In fact, however, the phase
space decomposition is necessary in order to derive a theory that allows
for a well-behaved perturbation theory. This can be explained most easily
by using the disorder term ${\cal A}_{\rm dis}$, Eqs.\ (\ref{eqs:2.10}),
as an example. As explained in I, the two contributions
${\cal A}_{\rm dis}^{(1,2)}$ result from a phase space decomposition analogous
to the one performed on the interaction, and the matrix $Q({\bf x})$ is
therefore to be understood as containing only Fourier components with
wavenumbers $\vert{\bf k}\vert < \lambda$. Now suppose we had not
performed the decomposition. Then ${\cal A}_{\rm dis}$ would consist of
${\cal A}_{\rm dis}^{(2)}$ only, with $\tau_{\rm rel}$ replaced by
$2\tau_{\rm rel}$, and $Q({\bf x})$ containing all Fourier components.
The saddle-point Green function for this action would then contain a
disorder part of the self-energy that is half the Born value. In a 
perturbation expansion in powers of the disorder, higher orders would
then have to make up for the missing factors of 2 at {\em zeroth} order,
i.e. the perturbation expansion would be singular. This is precisely
what happens in standard many-body perturbation theory,\cite{AGD} where
singular integrals make it impossible to easily determine the order of a
contribution from its diagrammatic structure, and infinite resummations
are in general necessary to obtain all contributions of a given order.
In contrast, perturbation theory for the nonlinear $\sigma$ model is
much better behaved, with the number of loops determining the order to
which a given diagram contributes. This is a consequence of a judicious
choice of the starting point for the loop expansion, which in turn
depends crucially on the phase space decomposition. The fact that the
theory depends on a cutoff is the price paid for the controlled nature
of perturbation expansions.

Similarly, the phase space decomposition performed on the interaction
part of the action allows us to perturbatively control the complicated
contribution $W_{\rm int}^{(2)}$ to the Ward identity, Eq.\ (\ref{eq:3.11}).
This guarantees that the nonlinear $\sigma$ model correctly reproduces
perturbation theory in the interaction to first order, as is well known
from comparing results obtained by either method.\cite{CDLM,Finkelstein,R}

Conversely, the above discussion makes it clear 
that the nonlinear $\sigma$ model,
owing to its derivation, is perturbative in nature with respect to the
electron-electron interaction. Indeed, its restriction in that respect is
more serious than with respect to disorder: Since the loop expansion is an
expansion in powers of the disorder, going to higher order in perturbation
theory will always include higher order disorder effects. With respect to
the interaction, the analogous statement is not true, since some effects of
higher order in the interaction are left out of the model,
although the loop expansion resums cetain classes of interaction terms
to all orders. Of course, a
complete renormalization of the model would in principle supply all of
the effects that might have been left out of the bare model, but this
will in general not be captured by the standard perturbative renormalization
based on low orders in the loop expansion. This is a restriction that is
important to keep in mind in the context of discussions about possible
exotic effects of a strong effective interaction, like e.g. a metallic
non-Fermi liquid ground state. It may also be relevant for understanding
the observation\cite{rf} that renormalization group calculations based
on the $\sigma$ model in high dimensions ($d>6$) reveal relevant terms of a
structure that is not seen in low-order $2+\epsilon$ expansions.
Physically, the `standard' generalized $\sigma$ model approach is valid if the 
physics one is interested in is determined by the two-particle diffusive modes.
If, for example, 
there were also soft single-particle excitations, then in general the 
coefficients in the $\sigma$ model would be singular and this approach would 
break down.
 
Finally, we note that already in the Gaussian approximation the perturbative 
nature of the $\sigma$ model approach with respect to interactions is apparent. 
In I we pointed out that before the massive modes were integrated out, the
Gaussian field theory explicitly contained the Stoner theory for 
ferromagnetism. However, after the $\sigma$ model approximation was made, 
the interaction terms that lead to the Stoner theory were absent.

\subsection{Screening, and the disordered Fermi-liquid fixed point}
\label{subsec:V.B}

A characteristic feature of the long-ranged Coulomb interaction is that
it leads to the system being incompressible: The wavenumber depended
thermodynamic derivative $(\partial n/\partial\mu)({\bf k}) =
\chi({\bf k},\Omega = 0)$, which is proportional to the compressibility,
is for small wavenumbers given by
\begin{equation}
\left(\frac{\partial n}{\partial\mu}\right)({\bf k}) = 
   \left(\frac{\partial n}{\partial\mu}\right)_{\rm sc}\,
        \frac{{\bf k}^2}{{\bf k}^2 + \kappa^2}\quad,
\label{eq:5.1}
\end{equation}
with $\kappa$ the screening wavenumber, and 
$(\partial n/\partial\mu)_{\rm sc} = \chi_{\rm sc}({\bf k}\rightarrow 0,
\Omega=0)$ the screened density susceptibility, which is a nonzero number.
This structure follows from, and is controlled by, the Ward identity,
as can be seen from Sec.\ \ref{subsec:III.B} above. It is instructive to
check explicitly that the nonlinear $\sigma$ model respects the
compressibility sum rule, Eq.\ (\ref{eq:5.1}). Within the framework of
the $\sigma$ model, one has\cite{R}
\begin{equation}
\left(\frac{\partial n}{\partial\mu}\right)({\bf k}) = 
    \frac{\pi}{8}\,\left(H + K_s({\bf k})\right)\quad,
\label{eq:5.2}
\end{equation}
with $H$ as defined after Eq.\ (\ref{eq:4.7}), and 
$K_s({\bf k}) = -\pi\Gamma^{(s)}({\bf k})/8$. We see that the $\sigma$
model indeed respects the compressibility sum rule, with the Gaussian
approximation for the screened compressibility,
$(\partial n/\partial\mu)_{\rm sc}\approx N_F$. Notice that in the model
originally proposed by Finkel'stein,\cite{Finkelstein} Fermi liquid
corrections had been put in to make $(\partial n/\partial\mu)_{\rm sc}$
the exact screened density susceptibility. These are missing here, since
we have integrated out the massive modes, which account
for the screening, in a Gaussian approximation. We emphasize that, while
it is of course always possible to put in Fermi liquid corrections by
hand, nothing is really gained by doing so: Such a procedure just amounts
to a partial resummation of some terms that are of higher order in the
interaction, which does not change the fact that the effective theory
has a perturbative character with respect to the interaction, as was
discussed in Sec.\ \ref{subsec:V.A} above. Furthermore, the point of any
effective theory is that it correctly captures the {\em structure} of the
full theory, while the coefficients can be represented by some approximation
in the bare theory. Upon renormalizing the bare effective theory, 
the coefficients will get renormalized by fluctuation effects.

\subsection{Renormalization group properties of the effective field theory}
\label{subsec:V.3}

We finally mention that the renormalization group properties of the
$\sigma$ model, Eq.\ (\ref{eq:4.7}), are well known. The theory possesses
a critical fixed point that describes an Anderson-Mott metal-insulator
transition.\cite{Finkelstein2,R} Also, due to the short-ranged nature
of the effective, screened, interaction (Eq.\ (\ref{eq:4.6})), the
discussion of the stable Fermi-liquid fixed point given in Sec. III.B.2 of
I still applies. The Fermi-liquid ground state is stable for $d>2$ in
the presence of quenched disorder, and for $d>1$ in the clean limit.
The corrections to scaling yield the weak-localization nonanalyticities
and their clean counterparts as discussed in I, modified by the
log-squared singularities in the density of states and the sound
attenuation that are induced by the Coulomb interaction\cite{AAKL,R}
as mentioned after Eq.\ (\ref{eq:3.10}) above. We emphasize again,
however, that due to the perturbative nature of the effective field
theory, our considerations do in no sense constitute a proof that the
Fermi-liquid fixed point will be stable for arbitrary strengths of the
bare interaction constant. What we have shown is that the fixed point
is perturbatively stable for weak interactions.
It is easy to see that an
interaction that is of longer range than the Coulombic one,
$v({\bf x}) \sim 1/\vert{\bf x}\vert^{\alpha}$ with $\alpha <1$, will
destroy the screening process, and hence lead to a relevant operator
that destroys the Fermi liquid fixed point, at least close to its lower
critical dimension. If an effective interaction of such long range
were generated by the renormalization group acting on the $\sigma$ model,
or if it were present in a the bare action for a different effective
theory that is not subject to the perturbative restriction of weak
interactions, then this could lead to a non-Fermi liquid ground state.
These points may be important in the context of the
ongoing discussion about possible non-Fermi liquid ground states in
$2$-$d$ (clean) electron systems.\cite{NFL}

\acknowledgments
This work was supported by the NSF under Grant Nos. DMR--95--10185,
DMR--96--32978, and DMR--98--70597, and by the Deutsche Forschungsgemeinschaft. 
Part of this work was performed at the Aspen Center for 
Physics. We thank the Center for its hospitality, and E. Abrahams for
a discussion concerning the perturbative nature of the $\sigma$ model.

\appendix
\section{Properties of the $Q$-matrices}
\label{app:A}

Here we list again the symmetry properties of the $Q$-matrices that
were defined in I, and add some additional remarks.

$Q$ is self-adjoint under the adjoint operation $Q^+ = C^T\,Q^T\,C$, with
$C_{nm}^{\alpha\beta} = \delta_{nm}\,\delta_{\alpha\beta}\,i\tau_1
      \otimes s_2\ $,
\begin{mathletters}
\label{eqs:A.1}
\begin{equation}
Q = C^T\,Q^T\,C\quad.
\label{eq:A.1a}
\end{equation}
In addition, the hermitian conjugate $Q^{\dagger}$ of $Q$ is related to $Q$
by,\cite{ErratumFootnote}
\begin{equation}
Q^{\dagger} = \left(\tau_3\otimes s_0\right)\,\Gamma\,Q\,\Gamma^{-1}\,
               \left(\tau_3\otimes s_0\right)\quad,
\label{eq:A.1b}
\end{equation}
where the similarity transformation denoted by $\Gamma$ has the property
\begin{equation}
(\Gamma\,Q\,\Gamma^{-1})_{nm} = Q_{-n-1,-m-1}\quad.
\label{eq:A.1c}
\end{equation}
\end{mathletters}%
We now expand our matrix fields in the spin-quaternion basis defined after
Eq.\ (\ref{eq:2.8'}),
\begin{mathletters}
\label{eqs:A.2}
\begin{eqnarray}
Q_{12}({\bf x})&=&\sum_{r,i=0}^{3}\,\left(\tau_r\otimes s_i\right)
   \,{_r^iQ}_{12}({\bf x})\quad,
\label{eq:A.2a}\\
\tilde\Lambda_{12}({\bf x})&=&\sum_{r,i=0}^{3}\,\left(\tau_r\otimes s_i
   \right)\,{_r^i\tilde\Lambda}_{12}({\bf x})\quad.
\label{eq:A.2b}
\end{eqnarray}
\end{mathletters}%
where again $1\equiv (n_1,\alpha_1)$, etc. 
In this basis, we have the following symmetry properties,
\begin{mathletters}
\label{eqs:A.3}
\begin{eqnarray}
{^0_r Q}_{12}&=&(-)^r\,{^0_r Q}_{21}\quad,\quad (r=0,3)\quad,
\label{eq:A.3a}\\
{^i_r Q}_{12}&=&(-)^{r+1}\,{^i_r Q}_{21}\ ,\ (r=0,3;\ i=1,2,3)\quad,
\label{eq:A.3b}\\
{^0_r Q}_{12}&=&{^0_r Q}_{21}\quad,\quad (r=1,2)\quad,
\label{eq:A.3c}\\
{^i_r Q}_{12}&=&-{^i_r Q}_{21}\quad,\quad (r=1,2;\ i=1,2,3)\quad.
\label{eq:A.3d}
\end{eqnarray}
\end{mathletters}%
Together with the behavior under hermitian conjugation, Eq.\ (\ref{eq:A.1b}),
this further implies
\begin{equation}
{^i_r Q}_{12}^* = - {^i_r Q}_{-n_1-1,-n_2-1}^{\alpha_1\alpha_2}\quad,
\label{eq:A.4}
\end{equation}
for all $i$ and $r$.\cite{ErratumFootnote}
Analogous relations hold for $\tilde\Lambda$ by virtue of the linear
coupling between $Q$ and $\tilde\Lambda$.

We further note that as a result of Eqs.\ (\ref{eq:A.3d}) and (\ref{eq:A.4})
the matrix elements ${^{1,2,3}_{1,2} Q}_{11}$ in the particle-particle
spin-triplet channel are real (not imaginary, as erroneously stated in I).
As a result, the Gaussian theory (Eqs. (2.36) in I) is formally unstable
in that channel, and the formally diverging Gaussian integral needs to
be interpreted. We have ascertained that a rotation of the relevant 
$Q$-integration contour onto the imaginary axis, which
effectively makes ${^{1,2,3}_{1,2} Q}_{11}$ imaginary, provides an 
interpretation that guarantees agreement with the well-known results
of conventional perturbation theory.\cite{AAKL} Also note that within the
$\sigma$ model, the matrix elements of $Q$ in the particle-particle 
spin-triplet channel are imaginary.\cite{R} This follows from 
Eq.\ (\ref{eq:A.1a}) in conjunction with the hermiticity of $Q$ within the
$\sigma$ model, Eq.\ (\ref{eq:4.4a}).

\end{document}